\documentstyle[12pt]{article}
\newcommand{\LL}{{I\!\! L}}
\newcommand{\eq}[1]{(\ref{#1})}
\newcommand{\diff}{\partial}
\newcommand{\be}{\begin{eqnarray}}
\newcommand{\ee}{\end{eqnarray}}
\newcommand{\cD}{{\cal D}}
\newcommand{\Z}{{Z \!\!\! Z}}
\newcommand{\wg}{\wedge}
\newcommand{\expb}[1]{\exp\Bigl\{ #1 \Bigr\} }
\def\cN{{\cal N}}
\def\cL{{\cal L}}
\def\dd{{\rm d}}
\def\NP{{\it Nucl.~Phys.}~}
\def\PR{{\it Phys.~Rev.}~}
\def\PL{{\it Phys.~Lett.}~}
\def\PRL{{\it Phys.~Rev.~Lett.}~}

\title{
\vspace{-1cm}
\begin{flushright}
{\large ITEP-TH-5/98}
\end{flushright}
\vspace{1.5cm}
Dyon Condensation and Aharonov--Bohm Effect}

\author{E.T.~Akhmedov\thanks{e--mail: akhmedov@vxitep.itep.ru},
M.N.~Chernodub and M.I.~Polikarpov\\~\\
{\sl ITEP, B. Cheremushkinskaya, 25, Moscow, 117259, Russia}
}

\date{}

\begin{document}
\maketitle
\thispagestyle{empty}

\begin{abstract}
  We derive the string representation of the Abelian Higgs theory
in which dyons are condensed. It occurs that in such
representation the topological interaction exists in the expectation
value of the Wilson loop. Due to this interaction the dynamics
of the string spanned on the Wilson loop is non-trivial.

\end{abstract}

\vspace{1cm}

\baselineskip=15pt

The method of abelian projections \cite{tH81} is one of the
popular approaches to the confinement problem \cite{Simonov} in
non--abelian gauge theories. Numerous computer simulations of the
lattice gluodynamics in the abelian projection (see, {\it e.g.}
Refs. \cite{LatticeRev's}) show that the vacuum of
gluodynamics behaves as a dual superconductor \cite{tHMa76}. The key
role in the dual superconductor model of the QCD vacuum is played by
abelian monopoles \cite{tH81}. In the abelian projection quarks are
electrically charged particles, and if monopoles are condensed the dual
Abrikosov string carrying the electric flux is formed between quark and
antiquark. Due to a non-zero string tension the quarks are confined by the
linear potential.

The abelian monopole currents in gluodynamics are
correlated \cite{InstMonKucha} with (anti-)ins\-tan\-tons. For the
(anti-)self-dual fields the abelian monopoles become abelian
dyons \cite{BoSc96}.  Moreover, in the vacuum of lattice gluodynamics
the local correlator of the topological charge density and the
product of the electric and magnetic currents is
positive \cite{ChGuPo98}. This means that the abelian monopoles have
the electric charge. The sign of this electric charge
coincides with the sign of the product of the magnetic charge and the
topological charge density. Thus the infrared
properties of the QCD in the abelian projection can be described by
the Abelian Higgs model (AHM) in which dyons are condensed. The
electric charge of the dyons fluctuates\footnote{Note that according
to the Schwinger quantization rule the electric charge $e$ of the
dyon is not fixed while magnetic charge $g$ is quantized: $e_0 g \in
2 \pi \cN$, $e_0$ is an elementary electric charge of an external
electric particle, see eq.\eq{<Wl>}.}.

Note that there exists the model of the QCD vacuum \cite{Simonov} in
which the {\it nonabelian} dyons are responsible for the confinement.
The nonabelian dyons (as instantons) give rise to the abelian dyons
in the abelian projection.

   Below we study the properties of the Abrikosov-Nielsen-Olesen
(ANO) strings in the abelian model in which dyons are condensed.  We
consider the abelian dyons which have a constant electric charge.  This model
can be a zero approximation for the realistic effective model of the QCD
vacuum in which the electric charge of the condensed dyons fluctuates.

   We start with the following expression for the partition function
in the Euclidian space--time\footnote{The theory with $e=0$
(monopoles are condensed) has been investigated as an effective
abelian theory of QCD in Refs.~\cite{MaSuzSug}.}:
\be
  Z_{dyon} =
  \int \cD A_{\mu} \cD B_{\mu} \cD \Phi \expb{ - \int \dd^4 x \,
  \cL_{dyon}(A,B,\Phi)}\,, \label{dyonZ}
\ee
where the dyon Lagrangian is:
\be
\cL_{dyon}(A,B,\Phi) = \cL_{gauge}(A,B) +
\frac{1}{2} {|(\diff_{\mu} - ieA_{\mu} - i g B_{\mu})\Phi|}^2 +
\lambda (|\Phi|^2 - \eta^2)^2\,.  \label{Lagrangian-dyon}
\ee
The field $B_{\mu}$ is the magnetic gauge potential, which is dual to
the electric gauge potential $A_{\mu}$, and $\Phi$ is the dyon field
with the electric charge $e$ and magnetic charge $g$. It was shown in
\cite{Zw71}, that it is possible to write the lagrangian in which
both fields $A_\mu$ and $B_\mu$ are regular:
\be
\cL_{gauge}(A,B) =
\frac{1}{2} {[n\cdot (\diff \wg A)]}^2
+ {1 \over 2} {[n\cdot (\diff \wg B)]}^2 + \nonumber\\
+ \frac{i}{2} {[n\cdot (\diff \wg A)]}^\nu
{[n\cdot ^*(\diff \wg B)]}_\nu
- \frac{i}{2} {[n\cdot (\diff \wg B)]}^\nu
{[n\cdot ^*(\diff \wg A)]}_\nu\,,\nonumber
\ee
where $[a\cdot (b\wg c)]^\nu \equiv a_\mu (b^\mu c^\nu -b^\nu
c^\mu)$, $[a\cdot ^*(b\wg c)]^\nu \equiv a_\mu \epsilon ^{\mu
\nu \alpha \beta }(b_\alpha c_\beta )$ and $n_\mu$ is an arbitrary
unit four-vector, $n^2 = 1$.

The partition function \eq{dyonZ} can be represented as the
partition function of the AHM. The lagrangian $\cL_{gauge}$ is invariant
under the linear transformation of the fields $A$ and $B$ \cite{Zw71}:
\be
{A \choose B} \to {A' \choose B'} = {{\cos \upsilon ~-\sin\upsilon}
\choose {\sin\upsilon ~~~\cos\upsilon}} {A \choose B}\,, \label{TrTheta}
\ee
where $\upsilon$ is an arbitrary constant. Applying this
transformation with the parameter
\be
\upsilon = - \arctan {g \over e}\,, \label{upsilon}
\ee
to eqs.(\ref{dyonZ},\ref{Lagrangian-dyon}) and integrating over the
field $A'$ we get the partition function of the AHM \cite{Zw71}:
\be
Z_{dyon} \propto Z_{AHM} = \int \cD {B'}_{\mu} \cD \Phi \expb{
- \int \dd^4 x \cL_{AHM}(B',\Phi)}\,,
\nonumber \\
\cL_{AHM}(B',\Phi) = \frac{1}{4} {(\diff_{[\mu} B'_{\nu]})}^2 +
\frac{1}{2} {|(\diff_{\mu} - i {\tilde g} B'_{\mu})\Phi|}^2
+ \lambda (|\Phi|^2 - \eta^2)^2\,, \label{AHM}
\ee
the Higgs field $\Phi$ has the magnetic charge\footnote{We call
$B_{\mu}'$ as the dual gauge field (thus $\Phi$ carries magnetic
charge) since we consider \eq{AHM} as the abelian effective model of
the QCD vacuum. Really after the transformation \eq{TrTheta} this is
the matter of convention.} $\tilde g = \sqrt{e^2 + g^2}$.

Consider the quantum average of the Wilson loop in the dyon theory
\eq{dyonZ}:
\be
<W^C_e>_{dyon} = \frac{1}{Z_{dyon}} \cdot \int \cD A_{\mu}
\cD B_{\mu} \cD \Phi \expb{ - \int \dd^4 x \, \cL_{dyon}(A,B,\Phi)}
W^C_e(A)\,, \label{<Wl>} \\
W^C_e(A) = \expb{i e_0 \int \dd^4 x \, j_\mu A^\mu }\,,
\qquad j_\mu(x) = \oint\limits_C \dd {\tilde x}_\mu
\, \delta^{(4)}(x - {\tilde x}(\tau))\,,\nonumber
\ee
which creates the particle with the electric charge $e_0$ on the
world trajectory\footnote{This average corresponds to the quark
Wilson loop if we consider \eq{dyonZ} as an effective theory of QCD.}
$C$.

Applying the transformation \eq{TrTheta}, \eq{upsilon} to the
quantum average \eq{<Wl>} and integrating over the field $A'_{\mu}$
we get:
\be
  <W^C_e>_{dyon} = {<K^C_{(q_e,q_m)}>}_{AHM}\,,
  \label{<Wl2>}
\ee
where the expectation value in the {\it r.h.s.}
of this equation is calculated in the AHM with the
lagrangian \eq{AHM}.  The operator $K$ is the product of the
t'Hooft loop \cite{tHo78} $H^C$ and the Wilson loop $W^C$:
\be
K^C_{(q_e,q_m)}
(B') = H^C_{q_e} (B') \cdot W^C_{q_m} (B')\,, \qquad q_e =
\frac{e_0 g}{\tilde g } \,, \qquad q_m = \frac{e_0 e}{\tilde g } \,.
\label{charges}
\ee
The operator $H^C_{q_e}$ is defined as follows:
\be
H^C_{q_e}(B') = \expb{ -
\frac{1}{4} \int \dd^4 x \left[  {(\diff_{[\mu} B'_{\nu]} - q_e \cdot
  \frac{1}{2} \varepsilon_{\mu \nu \alpha \beta} G^C_{\alpha
  \beta})}^2 - {(\diff_{[\mu} B'_{\nu]})}^2 \right]}\,, \label{tH}
\ee
where the tensor $G^C_{\mu\nu} = {(n \cdot \diff)}^{-1}
j_{[\mu} n_{\nu]}$ satisfies the relation $\diff_\nu
G^C_{\mu\nu} = j_\mu$. The tensor $F^d_{\mu\nu} = q_e
G^C_{\mu\nu}$ plays a role of the dual field strength tensor:
$\diff_\nu F^d_{\mu\nu} = q_e j_\mu$. In the string representation of
the AHM \cite{AkChPoZu96} the operator $H^C_{q_e}$ creates the
string spanned on the loop $C$, this string carries the flux
$q_e$.

The product $K^C$ of the operators $H^C$ and $W^C$ creates the dyon
loop with electric charge $q_e$ and magnetic charge $q_m$ on the
world trajectory $C$ in the AHM \eq{AHM}.

Now we discuss the string representation for the AHM \eq{AHM}
\cite{AkChPoZu96,PoWiZu93}.  In the center of the ANO strings the
field $\Phi = |\Phi| e^{i \theta}$ vanishes, $Im \Phi = Re \Phi = 0$,
and the phase $\theta$ is singular on the two dimensional surfaces,
which are world--sheets of the ANO strings.  The measure of the
integration over the fields $\Phi$ can be rewritten as follows:  $\cD
\Phi = const.\, \cD |\Phi|^2 \, \cD \theta$. $\int \cD
\theta$ contains the integration over functions which are singular on
two--dimensional manifolds, and we subdivide $\theta$ into the
regular $\theta^r$ and the singular $\theta^s$ parts:  $\theta =
\theta^r + \theta^s$, here $\theta^s$ is defined by:
\be
\diff_{[\mu,} \diff_{\nu]} \theta^s (x, \tilde x) = 2 \pi
\epsilon_{\mu \nu \alpha \beta} \Sigma_{\alpha \beta}(x, \tilde x),
\nonumber \\ \Sigma_{\alpha \beta}(x, \tilde x) =
\int\limits_{\Sigma} \dd^2 \sigma \, \epsilon^{ab} \diff_a\tilde
x_{\alpha}\diff_b\tilde x_{\beta}\, \delta^{(4)} [x - \tilde
x(\sigma)]\,, \quad \diff_a = \frac{\diff}{\diff\sigma^a}
\label{sigma}
\ee
the vector function $\tilde x_{\mu}$ is the position of the string,
$\Sigma$ is the collection of all closed surfaces, $\sigma =
(\sigma_1,\sigma_2)$ is the parametrization of the string surface;
the measure $\cD\theta$ can be decomposed as follows:  $\cD\theta =
\cD\theta^r \, \cD\theta^s$.

For simplicity we consider below the London limit of the AHM
($\lambda\to\infty$).  In this limit the radial part of the field
$\Phi$ is fixed everywhere except for the centers of the ANO
strings.  All expression below can be generalized to the
case of an arbitrary $\lambda$; this leads to an additional
functional integral over the radial part $|\Phi|$.

Performing the transformations as in
Refs.\cite{PoWiZu93,AkChPoZu96} we get the following string theory
for the quantum average \eq{<Wl>} of the Wilson loop:
\be
<W^C_e>_{dyon} = \frac{1}{Z_{str}}\int [\cD \tilde x] \cdot J(\tilde
 x) \cdot \exp\Biggl\{ - \int \dd^4 x \int \dd^4 y \Biggl [
 \nonumber\\ \frac{q^2_m}{2} j_\mu(x) D_m(x - y)j_\mu(y) + \pi i
 \zeta \cdot j_\mu(x) D_m(x - y) \diff_{\nu} \epsilon_{\mu \nu \alpha
 \beta} \left(\Sigma_{\alpha \beta}(y) + \cN G^C_{\mu \nu}(y)\right)
 \nonumber\\ + \pi^2 \eta^2 \left(\Sigma_{\mu\nu}(x) + \cN G^C_{\mu
 \nu}(x)\right) D_m(x - y) \left(\Sigma_{\mu \nu}(y) + \cN G^C_{\mu
 \nu}(y)\right) \Biggr] + 2\pi i \zeta \LL(\Sigma,C) \Biggr\}\,,
\label{<Wl>Str}
\ee
where
\be
   \cN = \frac{e_0 g}{2 \pi}\,, \qquad
   \zeta = \frac{e_0 e}{{\tilde g}^2}
   = \frac{e_0 e}{e^2 + g^2}\,,
   \label{N&zeta}
\ee
$D_m(x)$ is the scalar Yukawa propagator, $(\Delta + m^2)
D_m(x) = \delta^{(4)}(x)$, and $m^2 = 2 {\tilde g}^2 \eta^2$ is the
mass of the dual gauge boson ($B'$).

The measure $[\cD\tilde x_{\mu}]$ assumes both integration over all possible
positions and summation over all topologies of the string's world--sheets
$\Sigma$; $J(\tilde x)$ is the Jacobian of the transformation from the field
$\theta^s$ to the string position $\tilde x_{\mu}$.  The Jacobian $J(\tilde
x)$ was estimated in \cite{AkChPoZu96} for string with spherical or disc
topology.

   First three terms in the exponent in eq. \eq{<Wl>Str} describe the
short-range interaction and the self--interaction of the ANO strings
and dyon--anti-dyon pair through the exchange of the massive gauge
boson. The constant $\cN$ which appears in these terms has a physical
meaning. It is equal to the number of the elementary fluxes in the
string which connects the dyon--anti-dyon pair introduced by the
operator $K$, \eq{charges}.  By definition, $\cN = {q_e \over
\Psi_0}$, where $q_e$ is equal to the total electric flux from the
dyon and $\Psi_0 = \frac{2 \pi}{{\tilde g}}$ is the flux carried by
the elementary string in the AHM \eq{AHM}.  Since this number of the
elementary fluxes $\cN$ must be integer, we get the charge
quantization rule:  $e_0 g \in 2 \pi \cN$, $\cN \in \Z$ \cite{Zw71}.

The last term in eq.\eq{<Wl>Str},
\be
 \LL(\Sigma,C) = \frac{1}{4 \pi^2}
 \int \dd^4 x \int \dd^4 y\, \epsilon_{\mu\nu\alpha\beta}\,
 \Sigma_{\mu\nu}(x)\, j_{\alpha}(y)\,\frac{{(x-y)}_{\beta}
 }{{|x-y|}^4} \nonumber
\ee
is the linking number of the string world sheet $\Sigma$ and the
trajectory $C$ of the dyon. This formula represents
the long--range interaction which describes the {\it dual}
four--dimensional analogue \cite{ABE} of the dual Aharonov--Bohm
effect: strings correspond to electric solenoids which scatter
magnetic charges of abelian dyons. This linking number term is
important for the infrared properties of the theory since it may
induce an additional long range potential between quark and
anti-quark \cite{ABPotential}. It also leads to non-trivial commutation
relations between different operators in the theory \cite{AkChPoZu96}.

  The authors are gratefull to F.V.~Gubarev, Yu.A.~Simonov,
T.~Suzuki and A.V.~Zaharov for useful discussions. This work was
supported by the grants INTAS-96-370, INTAS-RFBR-95-0681,
RFBR-96-02-17230a and RFBR-96-15-96740.


\begin{thebibliography}{50}

\bibitem{tH81} G.~'t~Hooft, \NP {\bf B190} [FS3], 455 (1981).

\bibitem{Simonov} Yu.A.~Simonov, {\it Phys.Usp.} {\bf 39} (1996) 313.

\bibitem{LatticeRev's} T.~Suzuki, \NP {\bf B} {\it
(Proc.~Suppl.)} {\bf 30} (1993) 176; M.~I.~Polikarpov,
\NP {\bf B} {\it (Proc.~Suppl.)} {\bf 53} (1997) 134;
M.N.Chernodub and M.I.Polikarpov, {\it preprint ITEP-TH-55/97}, {\tt
hep-th/9710205}.

\bibitem{tHMa76}
G.~{'t Hooft}, in `High Energy Physics', ed. M.~Zichichi (Editrice
Compositori, Bolognia, 1976); S.~Mandelstam, {\it Phys.Rep.} {\bf
23C} (1976) 245.

\bibitem{InstMonKucha}
O.Miyamura and S.Origuchi, RCNP Confinement 1995, Osaka, Japan, March
22-26, 1995, p.137; M.N.Chernodub, F.V.Gubarev, {\it JETP Lett.} {\bf
62} (1995) 100; A.Hart and M.Teper, \PL {\bf 371} (1996) 261;
S.Thurner {\it et al.}, \PR {\bf D54} (1996) 3457; R.C.Brower, K.N.
Orginos, Chung-I Tan, \PR {\bf D55} (1997) 6313; M.Fukushima {\it et
al.}, \PL {\bf B399} (1997) 141.

\bibitem{BoSc96}
G.~Schierholz, RCNP Confinement 1995, Osaka, Japan, March
22-26, 1995, p.96, {\tt hep-lat/9506033}; V.Bornyakov, G.Schierholz,
\PL {\bf 384} (1996) 190.

\bibitem{ChGuPo98} M.N.~Chernodub, F.V.~Gubarev and M.I.~Polikarpov,
{\it preprint ITEP-TH-44/97}, {\tt hep-lat/9709039};
{\it preprint ITEP-TH-70/97}, {\tt hep-lat/9801010}.

\bibitem{ABPotential} F.A.~Bais, A.~Morozov, M.~de Wild~Propitius,
\PRL {\bf 71} (1993) 2383; M.N.~Chernodub,
F.V.~Gubarev and M.I.~Polikarpov, {\tt hep-lat/9704021}, to be
published in \PL {\bf B}; {\it Nucl.Phys.Proc.Suppl.}
{\bf 53} (1997) 581; {\tt hep-lat/9607045}.

\bibitem{MaSuzSug} S.~Maedan and T.~Suzuki, {\it Prog.~Theor.~Phys.}
{\bf 81} (1989) 229; H.~Suganuma, S.~Sasaki and H.~Toki, \NP {\bf
B435} (1995) 207.

\bibitem{Zw71} D.~Zwanziger, \PR {\bf D3} (1971) 880.

\bibitem{tHo78} G. 't Hooft, \NP {\bf B138} (1978) 1;
\NP {\bf B153} (1979) 141.

\bibitem{AkChPoZu96} E.~Akhmedov {\it et. all}, \PR
{\bf D53} (1996) 2087.

\bibitem{PoWiZu93} M.I.~Polikarpov, U.-J.~Wiese and M.A.~Zubkov
\PL {\bf 309B} (1993) 133; P.~Orland, \NP {\bf B428}, (1994)
221; M.~Sato and S.~Yahikozawa, \NP {\bf B436} (1995) 100.
E.T.~Akhmedov, {\it JETP Lett.} {\bf 64} (1996) 82.

\bibitem{ABE}
M.G.~Alford and F.~Wilczek, \PRL {\bf 62} (1989) 1071;
M.G.~Alford, J.~March--Russel and F.~Wilczek, \NP
{\bf B337} (1990) 695;
J.~Preskill and L.M.~Krauss, \NP {\bf B341} (1990) 50;
L.M.~Kraus and F.~Wilczek, \PRL {\bf 62} (1989) 1221.

\end{thebibliography}
\end{document}